%% file: paper_draft.tex
\documentclass[conference]{IEEEtran}

\usepackage{graphicx} 
\usepackage{xcolor, bm, amsmath}
\usepackage{amsfonts, amssymb}
\usepackage{mathtools}
\usepackage{enumitem}
\usepackage{comment}
\usepackage{lipsum}
\usepackage{soul,cite}
\usepackage{balance}
\usepackage[letterpaper,top=2cm,bottom=2.69cm,left=1.71cm,right=1.71cm]{geometry}
   \usepackage{flushend}
\usepackage{stackengine}

\usepackage[acronym,style=alttree,nomain,shortcuts,nonumberlist,nogroupskip,nopostdot]{glossaries}
 
\input{acronyms}

\newcommand{\RN}[1]{\textup{\uppercase\expandafter{\romannumeral#1}}}

\IEEEoverridecommandlockouts
\begin{document}

\title{Jamming Detection in Cell-Free MIMO \\ with Dynamic Graphs}
\author{\IEEEauthorblockN{Ali Hossary, Laura Crosara, and Stefano Tomasin
\thanks{This work is supported by the project ISP5G+ ( CUP D33C22001300002), which is part of the SERICS program (PE00000014) under the NRRP MUR program funded by the EU-NGEU" and by the European Commission through the Horizon Europe/JU SNS project ROBUST-6G (Grant Agreement no. 101139068). 
}
}
\IEEEauthorblockA{Dept.\ of Information Engineering (DEI), University of Padova, Italy \\
email: \{ali.hossary@,
laura.crosara@, stefano.tomasin@\}unipd.it}
}
\date{}
\maketitle

\thispagestyle{empty}
 \pagestyle{empty}

\begin{abstract}
Jamming attacks pose a critical threat to wireless networks, particularly in cell-free massive MIMO systems, where distributed access points and user equipment (UE) create complex, time-varying topologies. This paper proposes a novel jamming detection framework leveraging dynamic graphs and graph convolution neural networks (GCN) to address this challenge. By modeling the network as a dynamic graph, we capture evolving communication links and detect jamming attacks as anomalies in the graph evolution. A GCN-Transformers-based model, trained with supervised learning, learns graph embeddings to identify malicious interference. Performance evaluation in simulated scenarios with moving UEs, varying jamming conditions and channel fadings, demonstrates the method’s effectiveness, which is assessed through accuracy and F1 score metrics, achieving promising results for effective jamming detection.
\end{abstract}

\begin{IEEEkeywords}
Jamming Detection, Dynamic Graphs, and Graph Neural Networks.
\end{IEEEkeywords}

\glsresetall

\section{Introduction}
\label{sec:intro}

Wireless communication increasingly adopts cell-free architectures to enhance connectivity and spectral efficiency. Cell-free \ac{mimo} relies on \acp{ap} that jointly serve \acp{ue} without predefined cell boundaries. This paradigm shift introduces new challenges related to network dynamics and security \cite{ngo2017cell}.

As reliance on wireless services continues to grow, security threats have become a major concern. Wireless networks, due to the shared nature of the radio spectrum, are particularly vulnerable to jamming \cite{pirayesh2022jamming}. 
In \ac{mimo} wireless networks, traditional jamming detection methods rely on statistical models, which struggle to adapt to the complexities of dynamic wireless environments \cite{chandola2009anomaly}. In contrast, \ac{dl} techniques can be applied using a data-driven approach \cite{zhao2021deep}. In \cite{akhlaghpasand2018jamming}, a jammer detection method for massive MIMO systems is proposed, utilizing unused pilots during the training phase, assuming that the jammer lacks prior knowledge of the pilot patterns. The base station detects the presence of a jammer by analyzing the received signal on these unused pilots and employing a \ac{glrt}. 
Recent advancements have introduced new solutions, including \acp{nn} for jamming detection \cite{lohan2024from}.  
\Ac{dl} approaches, such as \acp{cnn}, have been employed in \cite{li2022jamming,zhang2019deep} to analyze spectrogram images for jamming detection, outperforming conventional feature-based methods. Recent advances are tailored to the characteristics of 5G networks \cite{10694335, 10810672, 10615325, 10464951}. In \cite{8851633}, a low-overhead intermittent jamming detection scheme for IoT networks is proposed, leveraging anchor nodes along with signal strength and multipath profile features. Furthermore, federated learning has been investigated for distributed jamming detection in flying ad-hoc networks \cite{mowla2020federated}. However, all these solutions are agnostic of the network structures and are not suited for cell-free communications where synchronization is looser.

When users are mobile and channel conditions vary, modeling network behavior is crucial. \textit{Dynamic graphs} offer a powerful representation for the evolving topology of wireless networks \cite{skarding2021foundations}, where nodes correspond to APs and UEs, and edges represent communication links based on signal strength and interference levels.  
To process and analyze dynamic graphs data, \textit{\acp{gnn}} provides a powerful framework. Inspired by \acp{cnn}, \acp{gnn} are designed to operate on graph structures, enabling tasks such as node classification, link prediction, and other graph-related learning problems \cite{wu2021comprehensive}.

In this paper, we propose a novel framework to model cell-free massive MIMO communication, exploiting dynamic graphs to capture the time variability of the communication scenario. Then, we present a novel approach for jamming detection, leveraging dynamic graphs and \ac{gnn}-based architectures. Our approach identifies jamming attacks by learning latent representations of network states and monitoring deviations from expected patterns. We evaluate the proposed method using simulations that model mobility, connectivity, and interference scenarios, demonstrating its effectiveness.

The rest of this paper is organized as follows. 
Section~II presents the cell-free MIMO system model. Section~III presents the GNN-based jamming detection framework. Section~IV evaluates detection performance through simulations. Finally, Section~V draws the conclusions.

\section{System Model}
\label{sec:sysmod}

We consider a cell-free massive \ac{mimo} network~\cite{elhoushy2022cell} with $M$ \acp{ap} and $M$ \acp{ue}, focusing on the downlink transmission. 
Each \ac{ue} is equipped with a single antenna, and each \ac{ap} is equipped with $N_{\rm A}$ antennas. 
\Acp{ap} are static, while \acp{ue} are moving. 
We adopt a discrete-time model with sampling interval $T$, considering the network state at time instants $nT$, with $n \in {\mathbb Z}$. Each AP is associated with a single UE, and uses maximal ratio precoding to transmit data to its served UE, we may have more UEs than AE, but still at any given time only one UE is connected to each AP.
Moreover, we account for the presence of a jammer that aims at corrupting the communication between \acp{ap} and \acp{ue}. Each \ac{ap} is transmitting with unitary power to each \ac{ue}.

\paragraph*{Channel Model} Let $\bm{h}(k,m, n)$ denote the $N_{\rm A} \times 1$ vector of the narrowband baseband equivalent channel between the $k$-th \ac{ue} and $m$-th \ac{ap} at time $nT$. 
We consider a Rician fading channel; thus, the channel vector is modeled as 
\begin{equation}\label{Ricianmodel}
\bm{h}_{k,m}(n) = \beta \sigma_{k,m}(n) + \sqrt{1-\beta^2}\bm{g}_{k,m}(n),
\end{equation}
with $\beta = \sqrt\frac{K}{K+1}$ a constant (and $K$ is the Ricean K-factor) and $\bm{g}_{k,m}(n)$ being a $N_{\rm A} \times 1$ random matrix having i.i.d. zero-mean complex Gaussian entries. The variance of each entry of $\bm{g}_{k,m}(n)$ is determined by the path-loss model, which characterizes the received signal power as a function of the distance $d_{k,m}(n)$ between the $k$-th \ac{ue} and the $m$-th \ac{ap} at time $nT$, i.e.,
\begin{equation}
\sigma^2_{k,m}(n) = \frac{d_0^2}{d_{k,m}^2(n)},
\end{equation}
with $d_0=100\,$m representing the distance at which the channel has unitary variance. 
With $\beta = 1$ we obtain a deterministic model, while varying $\beta\in[0,1]$ we configure the randomness of the fading channel. We assume that reception is affected by \ac{awgn} with variance $\sigma^2$ per antenna. 

\paragraph*{Signal-to-noise-plus-interference Ratio} The transmitter applies maximal ratio (MR) precoding to steer the transmitted signal towards the intended user, and, in the absence of jamming, a connection is established from the \ac{ap} $m$ to the \ac{ue} $k$ at time $nT$ if  the \ac{sinr} 
\begin{equation}
\Gamma_{k,m}(n) = \frac{|| \bm{h}_{k,m}(n) ||^4}{\sigma^2 + \sum_{m' \neq m}  |\bm{h}^H_{k,m}(n) \bm{h}_{k,m'}(n) |^2},
\end{equation}
is above a threshols $\Gamma_0$, i.e., 
\begin{equation}\label{cond}
\Gamma_{k,m}(n) > \Gamma_0.
\end{equation}
Note that the formula includes the interference from other APs.

\paragraph*{Mobility Model} We consider a system with \acp{ue} and \acp{ap} ditributed within a square area of edge length $L$. The coordinates of each \ac{ap}, indexed by $m$, are positioned at
fixed locations that cover the area. At $n=0$, the \acp{ue} are uniformly distributed within the square $[0, L]$. The coordinates of the position of user $k$ at time $nT$ are 
\begin{equation}\label{eq:pos}
\begin{split}
x_k(n+1) &= x_k(n) + (v_{x, k} + w_{x,k}(n+1))T, \\ 
y_k(n+1) &= y_k(n) + (v_{y, k} + w_{y,k}(n+1))T,  
\end{split}
\end{equation}
where $v_{x, k}$ and $v_{y, k}$ are the reference velocities of user $k$, uniformly distributed in the interval $[0,v_\mathrm{max}]$. The terms $w_{x, k}(n+1)$ and $w_{y, k}(n+1)$ are zero-mean Gaussian components with variance $\sigma^2_{w}$. 
If a user reaches the boundary of the square, its position is reset to a new location, uniformly sampled within the square, and assigned a new reference velocity.  We assume that each user maintains a minimum distance $d_\mathrm{min}$ from any \ac{ap}. 

\subsection{User Assignment Rule}

We adopt the following rule for the assignment of UE to its serving AP. We proceed iteratively. We start with the full list of APs and UEs, and select the UE $k$ and AP $m$ that have the minimum distance among all pairs in the list. We assign UE $k$ to AP $m$, then we remove a couple of devices from the list. The next iteration identifies the next AP-UE couple among the non-assigned APs and UEs. 

Note that this procedure generates the assignment between APs and UEs, while an effective communication link (connection) between each couple is obtained only if condition \eqref{cond} is satisfied.

\subsection{Jammer Behavior} We consider the presence of a jammer that intermittently affects the communication between \acp{ue} and \acp{ap}. Time is divided into $F$ frames, each of duration $T_\mathrm{F}$. Within each frame, the jammer remains active for a duration $\tau \in [0, T_\mathrm{F}]$. The jammer is equipped with a single antenna since its target is to disrupt any communication around it. When the jammer is {\em active}, the resulting \ac{sinr} for a transmission from \ac{ap} $m$ to \ac{ue} $k$ at time $nT$ becomes 
\begin{equation}
\Gamma_{k,m}(n) = 
 \frac{|| \bm{h}_{k,m}(n) ||^4}{\sigma^2 + P_{J}+\sum_{m' \neq m} |\bm{h}^H_{k,m}(n) \bm{h}_{k,m'}(n) |^2},
\end{equation}
where $\sigma^2_{\rm J}$ is the jammer transmit power, $P_{J} = \sigma^2_{\rm J} |S_k(n)|^2$, and $S_k(n)$ is the complex scalar channel from the jammer to \ac{ue} $k$ at time $nT$, according to the Rician model \eqref{Ricianmodel}.

\section{Jamming Detection By Dynamic Graph}
\label{sec:dyngraph}

We model the cell-free massive MIMO network as a dynamic connection graph $\{G(n)\}$, where $G(n)$ is the connection graph at time $nT$ and $T$ is the sampling time of the graph representation. In particular, each graph $G(n)$ has $N=2M$ nodes (in the set $V(n)$), corresponding to both the \acp{ap} and the \acp{ue}. The edges (collected in the set $E(n)$) represent the connections between \acp{ap} and \acp{ue}. Specifically, an edge exists between \ac{ue} $k$ and \ac{ap} $m$ when \eqref{cond} is satisfied.
Each edge from AP $m$ to UE $k$ is labeled with the vector $\bm{w}_{k,m}(n) = [\alpha d_{k,m}, \zeta \gamma_{k,m}]$, where $\alpha$ and $\zeta$ are normalization factors that ensure proper scaling between distance and SINR values. The edge weights encode key connectivity metrics:
\begin{itemize}
    \item \textit{connection distance} $d_{k,m}(n)$, which defines the physical distance between an \ac{ap} and a \ac{ue},
    \item \textit{link quality} $\Gamma_{k,m}(n)$, quantified by the \ac{sinr}, captures the reliability and performance of the communication link.
\end{itemize}

\subsection{Jamming Detection Technique}
 
Graph neural networks (GNNs)
are neural models that capture the dependence of graphs via message passing between the nodes of graphs. In
recent years, variants of GNNs such as graph convolutional network (GCN), graph attention network (GAT), and graph recurrent network (GRN) have demonstrated ground-breaking performances on many deep learning tasks\cite{ZHOU202057}. We propose a novel jamming detection framework to identify jamming attacks in wireless networks, based on the dynamic graph representation. The architecture leverages the dynamic graph $\{G(n)\}$, graph convolution, and attention mechanisms to capture the distinctive patterns of connectivity disruptions caused by signal jammers. 

The proposed jamming detection system consists of:
\begin{enumerate}
    \item \textbf{Feature Extraction}, Each static graph $G(n)$ is constructed from the network topology and connectivity data between nodes.
    \item \textbf{Spatial processing module (GCN layer):} Utilizes two stacked Gated Graph Convolutional layers to process each network snapshot independently and extract meaningful node-level representations (embeddings).
    \item \textbf{Temporal processing module (Transformer layer):} Applies a multi-head self-attention mechanism across a sequence of graphs to detect temporal patterns that are indicative of jamming.
    \item \textbf{Classification module:} Outputs a binary decision indicating whether the input sequence contains a jamming attack.

\end{enumerate}

\begin{figure}
    \centering
    \includegraphics[width=\linewidth]{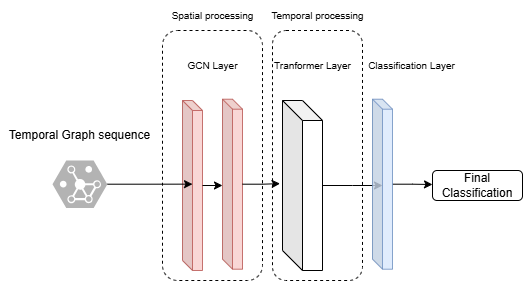}
    \caption{Architecture of the proposed jamming detection model. }
    \vspace{-0.3cm}
    \label{fig:arch}
\end{figure}

Fig. \ref{fig:arch} illustrates the overall architecture. The system processes sequences of $K$ network graphs $\mathcal G(t) = \{G(t), G(t+1), \ldots, G(t+K-1)\}$, where each sequence represents a specific network condition over time, to provide a binary decision on whether jamming activity is present within the sequence. 

The system processes a sequence of $N_{\text{steps}}$ consecutive network graphs:
\[
\mathcal{G}(n) = \{G(n), G(n+1), \ldots, G(n+N_{\text{steps}} - 1)\},
\]
where each $G(n)$ represents the state of the wireless network at time $nT$.

\subsubsection*{Feature Extraction and Graph Construction}

Each static graph $G(n)$ is constructed from the real-time network topology and connectivity data. From each graph, we extract the following features:

\begin{itemize}
    \item \textbf{Node-level features:}
    \begin{itemize}
        \item \textit{Degree centrality} $d_v(n)$: the number of connections of node $v$ at time $n$;
        \item \textit{Node type} $\tau_v \in \{0, 1\}$: where 0 denotes Access Points (APs) and 1 denotes User Equipments (UEs);
        \item \textit{Position coordinates} $(x_v(n), y_v(n))$: the physical location of node $v$ in 2D space.
    \end{itemize}

    \item \textbf{Edge-level features:}
    \begin{itemize}
        \item \textit{SINR} $\Gamma_{u,v}(n)$: the signal-to-interference-plus-noise ratio between nodes $u$ and $v$;
        \item \textit{Distance} $d_{u,v}(n)$: Euclidean distance between nodes $u$ and $v$, computed as:
        \[
        d_{u,v}(n) = \sqrt{(x_u(n) - x_v(n))^2 + (y_u(n) - y_v(n))^2}.
        \]
    \end{itemize}
\end{itemize}

These features are extracted from the dynamic graph object, which stores node types, positions, and connection weights between APs and UEs. After conducting ablation experiments by selectively removing features and measuring the resulting performance, we found the above features to be the most critical for detecting jamming events.

After experimenting with removing features and measuring performance degradation, the above-mentioned features are the most impactful for the jamming detection process.

\paragraph*{Spatial Processing Module} 
The extracted node and edge features are fed into a Graph Neural Network to compute node embeddings. These embeddings encode both the local structure (who a node is connected to) and attributes (such as position and type). Specifically, for each node $v$ at time $n$, we compute:
\[
h_v(n) = \text{GNN}(G(n), \bm{\xi}_v(n)),
\]
where $\bm{\xi}_v(n)$ is the feature vector of node $v$. The GCN aggregates information from neighboring nodes and edges, enabling each node to "learn" a summary of its local neighborhood and behavior.

\subsubsection*{Temporal Attention and Jamming Classification}

The sequence of node embeddings from each graph is passed to a Transformer layer. This layer uses temporal self-attention to identify patterns across time, specifically, it can emphasize graphs that exhibit abnormal behavior (such as sudden drops in SINR or rapid topology changes) and downweight normal periods. This is essential because jamming effects may not be constant but occur intermittently across the sequence.

\paragraph*{Classification Module} The final detection is performed by a single linear layer that classifies the aggregated representation.
The Transformer outputs a temporal representation $T_o(n)$, which is passed through a fully connected classification layer. The final output is the probability of jamming at the sequence level:

\begin{equation}p(n) = \text{Softmax}(\text{LayerNorm}(W \cdot T_o(n))).
\end{equation}

where LayerNorm denotes layer normalization, and $\bm{W}$ is the weight matrix of the classification layer.
The decision is based on whether the probability of the {\em jammer} class exceeds a fixed threshold.
This design allows the model to integrate spatial and temporal information effectively, improving robustness and interpretability in jamming detection

\subsection{Model Training}

The model is trained using the cross-entropy loss in a supervised manner using labeled datasets containing examples of nominal and jamming scenarios. During training, sequences of graph snapshots are presented to the model along with binary labels indicating the presence or absence of jamming activity. This supervised approach enables the model to learn discriminative patterns that distinguish normal network fluctuations from intentional jamming interference. The weights are optimized using the Adam optimizer, implementing early stopping when validation performance plateaus.

\section{Numerical Results}

\subsection{Dataset Generation}\label{sec:dataset}

To evaluate the proposed jamming detection approach, we generate a dataset of dynamic network graphs simulating wireless communications with and without jamming interference.

We consider a $L \times L$ area with $L =1\,$km, containing 5 fixed \acp{ap} and 10 mobile \ac{ue} nodes. The fixed APs are positioned at strategic locations covering the area: four at the corners, with coordinates $(0.2,0.2)$, $(0.8,0.2)$, $(0.2,0.8)$, and $(0.8,0.8)$, and one at the center $(0.5,0.5)$ (all in km unit). Mobile UEs move according to a controlled random walk model with velocity components drawn from a uniform distribution in $[-v_{\rm max}, v_{\rm max}]$, where $v_{\rm max} = 6~\text{km/h}$. We consider $T=1$~s and $T_F = 10$~s. Connectivity between an AP and UE is established when the \ac{sinr} exceeds the threshold $\Gamma_0 = 5~\text{dB}$. The noise power is $\sigma^2 = 0.001$. The jammer affects UEs within $0.35~\text{km}$ radius, and it is located in a different random position for each simulation. The number of network static graphs per sequence $\mathcal G(t)$ is  $N =80$.

We analyze two distinct scenarios. In the \textit{deterministic scenario}, we set $\beta = 1$, resulting in a fixed channel matrix $\bm{h}_{k,m}(n)$. In the \textit{fading scenario}, we set $\beta=0$, such that $\bm{h}_{k,m}(n)$ models a Rayleigh fading channel.

\subsection{GNN Implementation}
The architecture was implemented using PyTorch and PyTorch Geometric. We used a GCN layer for each snapshot of the dynamic graph that consists of 2 Gated Graph convolution layers with 64 hidden units. The Transformer encoder consists of 4 encoder layers, each with 16 attention heads, 64 hidden units, and a feed-forward dimension of 128. We use GELU activation in the feed-forward networks and apply layer normalization with batch-first processing. Since graph sequences have inherent temporal ordering, we add learned positional encodings to capture temporal relationships. A single linear layer with an intermediate dimension of 32 is used for binary classification. The model was trained for 30 epochs using the Adam optimizer with a learning rate of $1.2 \times 10^{-4}$, weight decay of $10^{-6}$, and batch size of 8. We applied a dropout of 0.03 in the Transformer layers and 0.05 overall to prevent overfitting.
The dataset has 2200 dynamic graphs for each scenario, training was performed on 70\% of the dataset, while 10\% of the dataset was used for validation and 20\% for testing.

\subsection{Performance Metrics}
Let TP be the number of True Positives, TN be the number of True Negatives, FP be the number of False Positives, and FN be the number of False Negatives. The accuracy is
\begin{equation}
a = \frac{\text{TP} + \text{TN}}{\text{TP} + \text{TN} + \text{FP} + \text{FN}}\,, 
\end{equation}

F1 score is
\begin{equation}
F_1 = \frac{2  \text{TP}}{2 \text{TP} + \text{FP} + \text{FN}}\,.
\end{equation}

\subsection{Simulation Results}

This section presents a comprehensive experimental evaluation of our dynamic graph-based jammer detection system under two primary training scenarios: (1) mixed-$\tau$ training using data from all jammer persistence patterns $\tau \in \{1,2,...,10\}$, and (2) $\tau=10$ specialist training using only continuous jammer scenarios. The parameter $\tau$ represents the jammer activation frequency within each temporal sequence, where $\tau=1$ indicates sporadic jamming (active for only 1 out of 10 timesteps), $\tau=5$ represents moderate persistence (active for 5 out of 10 timesteps), and $\tau=10$ denotes continuous jamming (active throughout the entire sequence). All experiments were conducted with 80-timestep sequences on cell-free MIMO networks, evaluating performance under both fading and non-fading channel conditions.

\subsection{$\tau=10$ Specialist Training Analysis}

\begin{figure}
    \centering
    \vspace{-0.18cm}
    \includegraphics[width=0.95\linewidth]{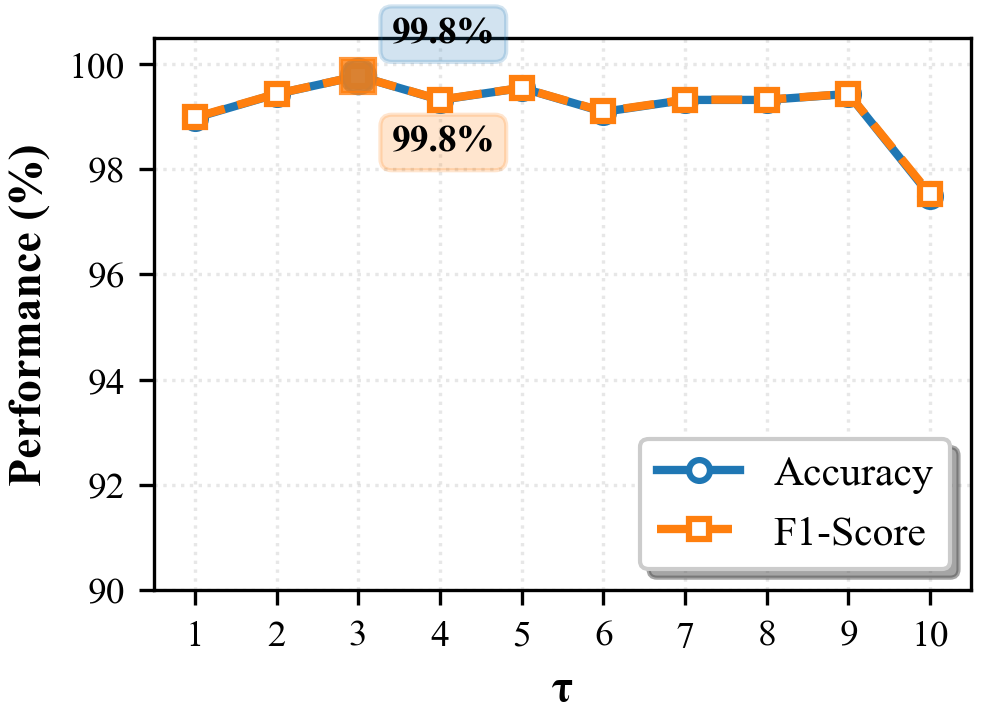}
    \caption{Accuracy and F1 score vs $\tau$, for the deterministic scenario. Training performed with a dataset having $\tau =10$.} 
    \label{fig:tau10_no_fading}
\end{figure}
The $\tau=10$ specialist results under non-fading conditions, shown in Fig.~\ref{fig:tau10_no_fading}, achieved accuracy consistently above 99\% across $\tau=1-9$, and F1-scores reaching 99.8\% at $\tau=3$. However, a notable performance degradation occurs at $\tau=10$, where accuracy drops to 97.1\% and F1-score to 97.4\%.
 This indicates that training exclusively on continuous jammer scenarios, counterintuitively, provides excellent generalization to sporadic and rhythmic jamming patterns under non-fading channels.

\begin{figure}
    \centering
    \includegraphics[width=0.95\linewidth]{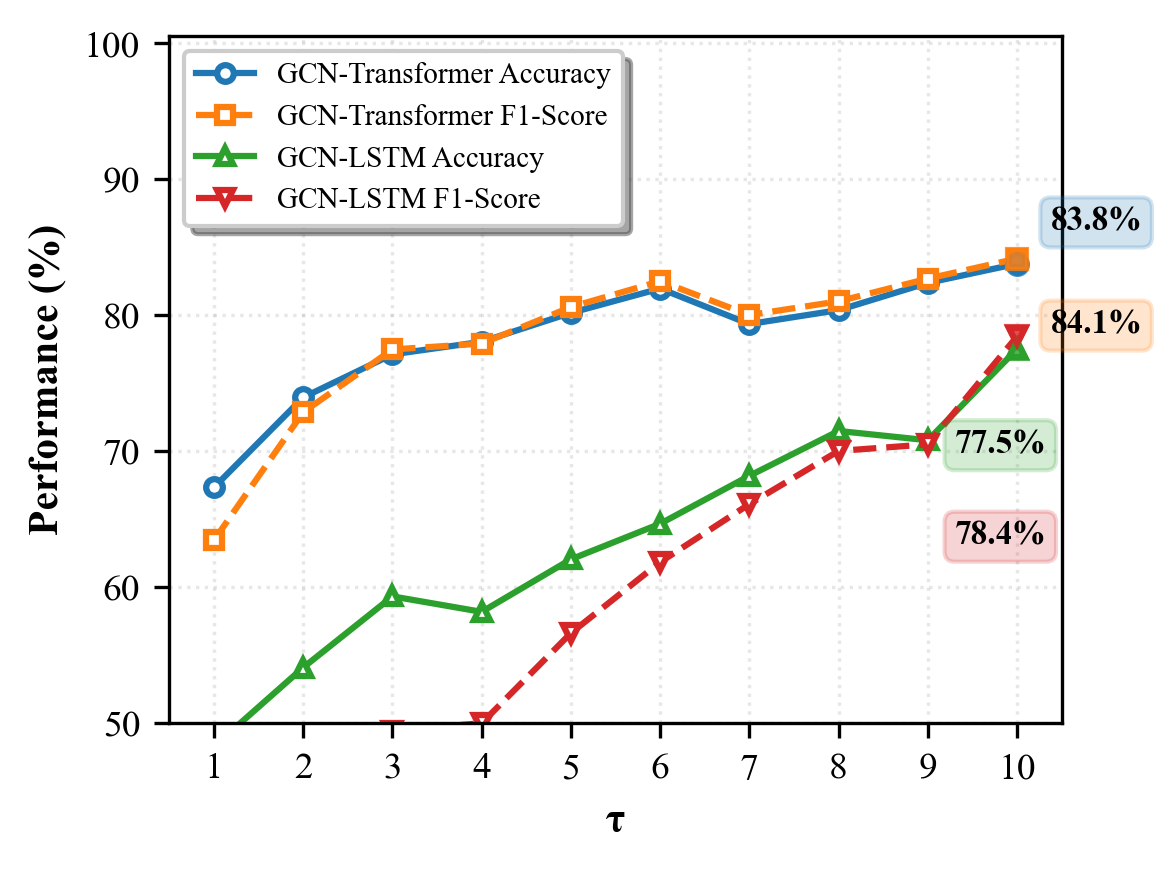}
    \caption{Accuracy and F1 score vs $\tau$, for the fading scenario. Training performed with a dataset having $\tau=10$.} 
    \vspace{-0.3cm}
    \label{fig:tau10_fading}
\end{figure}
In contrast, the fading scenario, shown in Fig.~\ref{fig:tau10_fading}, reveals the specialist's true generalization limitations and more pronounced performance variations. While maintaining strong overall performance (accuracy range: 67.2\%-83.8\%), the model shows increased sensitivity to jammer persistence patterns, however, a comparison has been done on the same dataset using the known Long Short Term Memory GCN (GCN-LSTM)\cite{García-Duarte_Cifuentes_Marulanda_2022} which combines the capabilities of LSTMs to extract temporal dependencies with the feature learning power of the GCN, and as the figure shows, our model performed better in all projected jamming behaviours. The performance progression from $\tau=1$ (67.2\% accuracy) to $\tau=8$ (83.8\% accuracy) demonstrates the model's adaptation to different temporal structures, with optimal detection occurring in the rhythmic jamming domain ($\tau=6-8$).
\subsection{Mixed-$\tau$ Training Performance}

\begin{figure}
    \centering
    \includegraphics[width=0.95\linewidth]{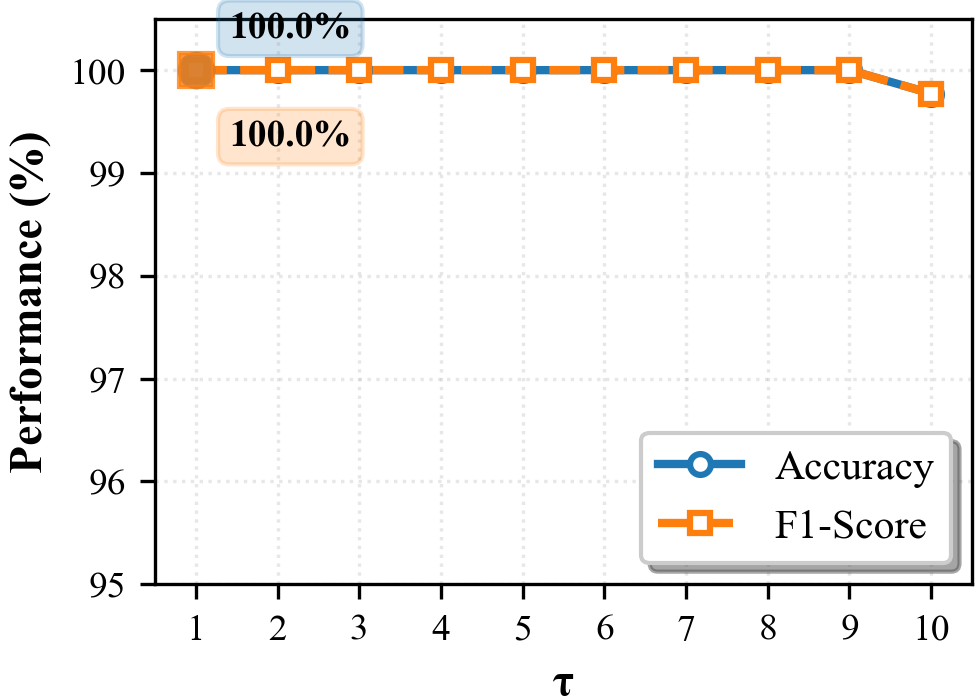}
    \vspace{-0.15cm}
    \caption{Accuracy and F1 score vs $\tau$, for the deterministic scenario. Training performed with a dataset having a mixture of attacks with different values of $\tau$.}
    \label{fig:mixed_tau_no_fading}
\end{figure}
Fig.~\ref{fig:mixed_tau_no_fading} shows the performance of our mixed-$\tau$ training approach under non-fading channel conditions. The model exhibit $100\%$ accuracy across $\tau=1-9$, with minimal degradation to 99.7\% at $\tau=10$.

\begin{figure}
    \centering
    \includegraphics[width=0.95\linewidth]{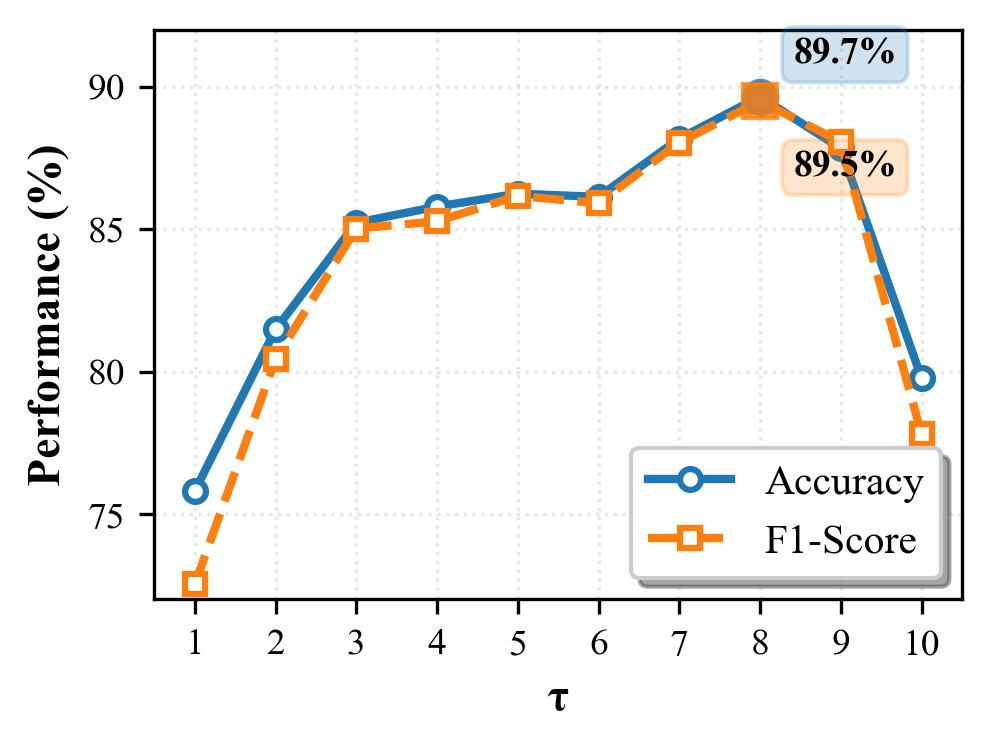}
    \caption{Accuracy and F1 score vs $\tau$, for the random fading scenario. Training performed with a dataset having a mixture of attacks with different values of $\tau$.}
    \label{fig:mixed_tau_fading}
\end{figure}
In the fading scenario, presented in Fig.~\ref{fig:mixed_tau_fading}, the obtained accuracy ranges from 75.6\% at $\tau=1$ to 89.7\% at $\tau=8$, before decreasing to 79.4\% at $\tau=10$. The monotonic improvement from $\tau=1$ to $\tau=8$ (73.2\% to 89.5\% F1-score) suggests that the model learns increasingly effective detection strategies as jammer persistence increases, until reaching the domain boundary at $\tau=9-10$.

\subsection{Channel Fading Effects on Detection Performance}

Comparing non-fading versus fading scenarios reveals significant differences in detection robustness. Under non-fading conditions, both training strategies achieve near-perfect performance across most $\tau$ values, suggesting that the absence of channel fading provides cleaner signal characteristics that enhance jammer detection reliability. The stable channel conditions appear to preserve jamming signatures without additional noise from natural channel variations.

Conversely, fading scenarios present more challenging detection environments, with performance variations of 10-15 percentage points across different $\tau$ values. This increased difficulty under fading channels indicates that channel-induced signal variations may mask jamming signatures, requiring more sophisticated detection algorithms to distinguish between fading-induced and jammer-induced signal degradations.

\subsection{Training Strategy Effectiveness Comparison}

The mixed-$\tau$ training approach demonstrates improved generalization capabilities and overall performance compared to the $\tau=10$ specialist across both channel conditions. Under non-fading conditions, mixed-$\tau$ training achieves near-perfect performance (more than 99\% accuracy) across the entire $\tau$ spectrum, while under fading conditions, it maintains reasonable performance levels (76-90\% range) with more graceful degradation patterns. In contrast, the $\tau=10$ specialist, despite showing perfect performance under non-fading conditions, exhibits significant generalization limitations under fading scenarios, with performance dropping as low as 67\% at $\tau=1$.

The mixed-$\tau$ approach's enhanced robustness across different channel conditions and jammer persistence patterns indicates that exposure to diverse jamming behaviors during training provides more generalizable feature representations. This finding supports the hypothesis that multi-domain training strategies are essential for robust jammer detection in dynamic wireless environments.

\subsection{Baseline Shift Problem in Persistent Jamming}

The performance degradation observed at $\tau=10$ across all experimental configurations can be attributed to the fundamental baseline shift problem in persistent jamming scenarios. When jammers operate continuously, cell-free MIMO networks undergo adaptive responses.
These network adaptations effectively establish a new operational baseline where continuous interference becomes the "normal" state.

\section{Conclusions}
\label{sec:concl}

This paper presented a comprehensive analysis of jammer detection in cell-free MIMO networks using dynamic graphs and specific graph neural network architecture, revealing insights into the effect of channel fading in the jamming detection process, and the multi-domain nature of temporal anomaly detection, in addition to this, our experimental evaluation across different jammer patterns ($\tau \in \{1,2,...,10\}$) demonstrated that mixed-$\tau$ training achieves enhanced generalization compared to specialist approaches, with performance exceeding 99\% under non-fading conditions and maintaining robustness above 75.6\% even in challenging fading scenarios, higher than existing known models. The comparative analysis between fading and non-fading channels revealed that stable channel conditions significantly enhance detection reliability, while channel fading introduces additional complexity that degrades performance by 10-15 percentage points across all $\tau$ values.

A nice finding of this work is the identification of the baseline shift problem in persistent jamming scenarios ($\tau=9-10$), where continuous jammer presence causes network adaptation responses that establish a new operational baseline, making traditional anomaly detection approaches ineffective. This phenomenon explains the characteristic performance degradation observed at high $\tau$ values across all experimental configurations, highlighting the need for detection strategies that can identify adaptation artifacts rather than direct interference signatures. The delineation of three distinct detection domains, namely sporadic ($\tau=1-3$), rhythmic ($\tau=4-8$), and persistent ($\tau=9-10$), provides a theoretical framework for developing domain-specific architectures that address the unique challenges of each jammer behavior pattern.

\bibliographystyle{IEEEtran}
\bibliography{biblio}

\end{document}

%% file: acronyms.tex
\newacronym{5gc}{5GC}{5G core network}
\newacronym{5gprs}{5G PRS}{5G positioning reference signal}

\newacronym{acas}{ACAS}{assisted commercial authentication service}
\newacronym{acf}{ACF}{auto-correlation function}
\newacronym{adc}{ADC}{analog to digital converter}
\newacronym{agc}{AGC}{automatic gain control}
\newacronym{ae}{AE}{autoencoder}
\newacronym{an}{AN}{artificial noise}
\newacronym{aoi}{AoI}{age of information}
\newacronym{ap}{AP}{access point}
\newacronym{arq}{ARQ}{automatic repeat request}
\newacronym{auv}{AUV}{autonomous underwater vehicle}
\newacronym{awgn}{AWGN}{additive white Gaussian noise}

\newacronym{bds}{BDS}{BeiDou navigation satellite system}
\newacronym{ber}{BER}{bit error rate}
\newacronym{bgd}{BGD}{broadcast group delay}
\newacronym{bilp}{BILP}{binary integer linear programming}
\newacronym{boc}{BOC}{binary offset carrier}
\newacronym{bp}{BP}{bounded PER}
\newacronym{bpsk}{BPSK}{binary phase shift keying}
\newacronym{bqp}{BQP}{bounded-error quantum polynomial}
\newacronym{bs}{BS}{base station}

\newacronym{ca}{C/A}{coarse/acquisition}
\newacronym{cas}{CAS}{commercial authentication service}
\newacronym{cboc}{CBOC}{composite binary offset carrier}
\newacronym{cdf}{CDF}{cumulative distribution function}
\newacronym{cdma}{CDMA}{code division multiple access}
\newacronym{chim}{CHIMERA}{chips-message robust authentication}
\newacronym{cldae}{CLDAE}{combined \ac{ld} and \ac{ae}}
\newacronym{cnn}{CNN}{convolutional neural network}
\newacronym{cots}{COTS}{commercial off-the-shelf}
\newacronym{cp}{CP}{cyclic prefix}
\newacronym{cpu}{CPU}{central processing unit}
\newacronym{crc}{CRC}{cyclic redundancy check}
\newacronym{cr}{CR}{challenge response}
\newacronym{crlb}{CRLB}{Cramér-Rao lower bound}
\newacronym{cr-pla}{CR-PLA}{challenge response \ac{pla}}
\newacronym{cs}{CS}{commercial service}
\newacronym{csi}{CSI}{channel state information}

\newacronym{det}{DET}{detection error tradeoff}
\newacronym{dft}{DFT}{discrete Fourier transform}
\newacronym{dl}{DL}{deep learning}
\newacronym{dll}{DLL}{delay-locked-loop}
\newacronym{dos}{DoS}{denial of service}
\newacronym{dsm}{DSM}{digital signature message}
\newacronym{dsp}{DSP}{digital signal processing}
\newacronym{dsss}{DS-SS}{direct sequence spread spectrum}

\newacronym{ecef}{ECEF}{earth-centered earth-fixed}
\newacronym{ekf}{EKF}{extended Kalman filter}
\newacronym{enu}{ENU}{East, North, Up}
\newacronym{ecs}{ECS}{encrypted code sequence}
\newacronym{esa}{ESA}{European Space Agency}
\newacronym{estec}{ESTEC}{European Space Research and Technology Centre}         
\newacronym{eu}{EU}{European Union}
\newacronym{euspa}{EUSPA}{European Union Agency for the Space Programme}

\newacronym{fa}{FA}{false alarm}
\newacronym{fcfs}{FCFS}{{first-come-first-served}}
\newacronym{fec}{FEC}{forward error correction}

\newacronym{gcn}{GCN}{graph convolutional network}
\newacronym{gnn}{GNN}{graph neural network}
\newacronym{gdop}{GDOP}{geometric dilution of precision}
\newacronym{geo}{GEO}{geostationary Earth orbit}
\newacronym{ggto}{GGTO}{GPS to Galileo time offset}
\newacronym{gmm}{GMM}{Gaussian mixture model}
\newacronym{glonass}{GLONASS}{global'naja navigacionnaja sputnikovaja sistema}
\newacronym{glrt}{GLRT}{generalized likelihood ratio test}
\newacronym{gnav}{G/NAV}{governmental navigation message}
\newacronym{gnss}{GNSS}{global navigation satellite system}
\newacronym{gps}{GPS}{global positioning system}
\newacronym{gsa}{GSA}{European GNSS Agency}
\newacronym{gsc}{GSC}{GNSS service center}
\newacronym{gst}{GST}{Galileo system time}

\newacronym{has}{HAS}{high accuracy service}
\newacronym{hpl}{HPL}{horizontal protection level}

\newacronym{icto}{ICTO}{inter-constellation time offset}
\newacronym{ici}{ICI}{inter carrier interference}
\newacronym{id}{ID}{identification number}
\newacronym{if}{IF}{intermediate frequency}
\newacronym{iid}{i.i.d.}{independent and identically distributed}
\newacronym{ilp}{ILP}{integer linear programming}
\newacronym{imu}{IMU}{inertial measurement unit}
\newacronym{inav}{I/NAV}{integrity navigation message}
\newacronym{iot}{IoT}{Internet of things}
\newacronym{ip}{IP}{internet protocol}
\newacronym{irs}{IRS}{intelligent reflecting surface}
\newacronym{iscb}{ISCB}{inter-system clock bias}
\newacronym{isi}{ISI}{inter-symbol interference}

\newacronym{kde}{KDE}{kernel density estimation}
\newacronym{kl}{KL}{Kullback–Leibler}

\newacronym{lan}{LAN}{local area network}
\newacronym{lc}{LC}{linear combination}
\newacronym{lcfs}{LCFS}{{last-come-first-served}}
\newacronym{ld}{LD}{local decision}
\newacronym{leo}{LEO}{low Earth orbit}
\newacronym{llh}{LLH}{latitude-longitude-height}
\newacronym{lmf}{LMF}{location management function}
\newacronym{lo}{LO}{local oscillator}
\newacronym{los}{LOS}{line-of-sight}
\newacronym{lpp}{LPP}{LTE positioning protocol}
\newacronym{lrt}{LRT}{likelihood ratio test}
\newacronym{ls}{LS}{least squares}
\newacronym{lssvm}{LS-SVM}{least squares support vector machines}

\newacronym{ma}{MA}{moving average}
\newacronym{mac}{MAC}{message authentication code}
\newacronym{map}{MAP}{maximum a posteriori}
\newacronym{marp}{MARP}{Maximization of the Average Number of Different Received Packets}
\newacronym{marpmc}{MARP-MC}{Maximization of the Average Number of Different Received Packets among Maximum Coverage Solutions}
\newacronym{meo}{MEO}{medium Earth orbit}
\newacronym{md}{MD}{missed detection}
\newacronym{mid}{MID}{max inter-group distance}
\newacronym{mimo}{MIMO}{multiple-input multiple-output}
\newacronym{milp}{MILP}{mixed-integer linear programming }
\newacronym{minmax}{MIN-MAX}{Minimization of the Maximum Latency}
\newacronym{mitm}{MITM}{man-in-the-middle}
\newacronym{miw}{MIW}{min inter-group weight}
\newacronym{ml}{ML}{machine learning}
\newacronym{mlp}{MLP}{multilayer perceptron}
\newacronym{mle}{MLE}{maximum likelihood estimate}
\newacronym{mp}{MP}{min-max PER}
\newacronym{moregossip}{MORE GOSSIP}{More GNSS Open Service Signal Integrity Protection and Authentication at the Physical Layer}

\newacronym{mse}{MSE}{mean squared error}
\newacronym{mmse}{MMSE}{minimum \ac{mse}}

\newacronym{nav}{NAV}{navigation message}
\newacronym{nco}{NCO}{numerically controlled oscillator}
\newacronym{ne}{NE}{Nash Equilibrium}
\newacronym{nlos}{NLOS}{non-line-of-sight}
\newacronym{nma}{NMA}{navigation message authentication}
\newacronym{nme}{NME}{Navigation Message Encryption}
\newacronym{nmea}{NMEA}{national marine electronics association}
\newacronym{nn}{NN}{neural network}
\newacronym{nr}{NR}{new radio}
\newacronym{ntp}{NTP}{network time protocol}
\newacronym{ntn}{NTN}{non-terrestrial network}

\newacronym{ocsvm}{OC-SVM}{one-class support vector machine}
\newacronym{ofdm}{OFDM}{orthogonal frequency division multiplexing}
\newacronym{os}{OS}{open service}
\newacronym{osnma}{OSNMA}{open service navigation message authentication}
\newacronym{otp}{OTP}{one-time pad}

\newacronym{patrol}{PATROL}{Position Authenticated Tachograph for OSNMA Launch}
\newacronym{pdop}{PDOP}{position dilution of precision}
\newacronym{pdf}{pdf}{probability density function}
\newacronym{per}{PER}{packet error rate}
\newacronym{pla}{PLA}{physical layer authentication}
\newacronym{pll}{PLL}{phase-locked-loop}
\newacronym{pls}{PLS}{physical-layer security}
\newacronym{pnt}{PNT}{position navigation and timing}
\newacronym{pntcrc}{PNT-CRC}{Position Navigation and Timing Cyber-Response Center}
\newacronym{ppp}{ppp}{precise point positioning}
\newacronym{ppm}{ppm}{parts per million}
\newacronym{pps}{PPS}{precise positioning service}
\newacronym{prb}{PRB}{physical resource block}
\newacronym{prn}{PRN}{pseudorandom noise}
\newacronym{prs}{PRS}{public regulated service}
\newacronym{psd}{PSD}{power spectral density}
\newacronym{ptp}{PTP}{precision time protocol}
\newacronym{pvt}{PVT}{position, velocity, and time}

\newacronym{raim}{RAIM}{receiver autonomous integrity monitoring}
\newacronym{re}{RE}{resource element}
\newacronym{recs}{RECS}{re-encrypted code sequence}
\newacronym{rnn}{RNN}{recurrent neural network}
\newacronym{relu}{ReLu}{rectified linear unit}
\newacronym{rf}{RF}{radio frequency}
\newacronym{rinex}{RINEX}{receiver independent exchange format}
\newacronym{rms}{RMS}{root mean square}
\newacronym{rmse}{RMSE}{root mean squared error}
\newacronym{rng}{RNG}{random number generator}
\newacronym{roc}{ROC}{receiver operating characteristic}
\newacronym{rs}{RS}{random scheduling}
\newacronym{rsa}{RSA}{Rivest–Shamir–Adleman}
\newacronym{rtt}{RTT}{round trip time}
\newacronym{rv}{r.v.}{random variable}

\newacronym{sar}{SAR}{search and rescue service}
\newacronym{sas}{SAS}{signal authentication service}
\newacronym{sbas}{SBAS}{satellite-based augmentation systems}
\newacronym{sbf}{SBF}{Septentrio binary format}
\newacronym{sce}{SCE}{spreading code encryption}
\newacronym{sca}{SCA}{spreading code authentication}
\newacronym{scer}{SCER}{security code estimation and replay}
\newacronym{sdr}{SDR}{software defined radio}
\newacronym{sinr}{SINR}{signal-to-interference-plus-noise ratio}
\newacronym{sis}{SIS}{signal in space}
\newacronym{siso}{SISO}{single-input single-output}
\newacronym{snr}{SNR}{signal-to-noise ratio}
\newacronym{sqm}{SQM}{signal quality monitoring}
\newacronym{ssa}{SSA}{single-sensor authentication}
\newacronym{ssp}{SSP}{sound speed profiles}
\newacronym{sssc}{SSSC}{spread spectrum security code}
\newacronym{sv}{SV}{space vehicle}
\newacronym{svm}{SVM}{support vector machines}

\newacronym{tcp}{TCP}{transmission control protocol}
\newacronym{tcxo}{TCXO}{temperature-controlled crystal oscillator}
\newacronym{tdma}{TDMA}{time-division multiple access}
\newacronym{tesla}{TESLA}{timed-efficient stream loss-tolerant authentication} 
\newacronym{tmboc}{TMBOC}{time-multiplexed BOC}
\newacronym{tta}{TTA}{time to authenticate}
\newacronym{tn}{TN}{technical note}
\newacronym{toa}{TOA}{time of arrival}

\newacronym{uav}{UAV}{unamed aerial vehicle}
\newacronym{ue}{UE}{user equipment}
\newacronym{ula}{ULA}{uniform linear array}
\newacronym{utc}{UTC}{coordinated universal time}
\newacronym{uere}{UERE}{user equivalent ranging error}
\newacronym{uls}{ULS}{uplink station}
\newacronym{uwac}{UWAC}{underwater acoustic channel}
\newacronym{uwan}{UWAN}{underwater acoustic network}

\newacronym{vctcxo}{VCTCXO}{voltage-controlled and temperature-controlled crystal oscillator}
\newacronym{val}{VAL}{vertical alert limit}
\newacronym{vel}{VEL}{vertical error level}
\newacronym{vir2em}{VIR2EM}{VIrtualization and Remotization for Resilient and Efficient Manufacturing}
\newacronym{vpl}{VPL}{vertical protection level}

\newacronym{wsn}{WSN}{wireless sensor network}